\documentclass[aps,floats,twocolumn,prl,nofootinbib]{revtex4-2}
\usepackage{amsfonts,amsmath,amssymb,ascmac,bm,amsthm}
\usepackage{fnpct} 
\usepackage{comment}
\usepackage{ifpdf}
\usepackage{graphicx}
\usepackage{slashed}
\usepackage{color}
\usepackage[mathscr]{eucal}
\usepackage[utf8]{inputenc}
\usepackage{physics}
\usepackage{cancel}
\usepackage{float}
\usepackage{soul}
\usepackage{booktabs}
\usepackage{simpler-wick}
\usepackage{hyperref}
\usepackage{tensor}
\usepackage{upgreek}
\usepackage{caption}
\usepackage{subfigure}
\usepackage{subcaption}
\usepackage{tikz}
\usepackage{appendix}

\newcommand\mi{\mathrm{i}}
\newcommand\me{\mathrm{e}}

\newcommand{\dif}{\mathrm{d}}

\newcommand{\mPsi}{{\mit \Psi}}
\newcommand{\mTheta}{{\mit \Theta}}

\begin{document}

\title{Parity violating spectral dynamics of black holes in dynamical Chern-Simons gravity}

\author{Han-Wen Hu$^{1,2}$}
\email{huhanwen@itp.ac.cn}
\author{Chen Lan$^{3}$}
\email{stlanchen@126.com}
\author{Zong-Kuan Guo$^{1,2,4}$}
\email{guozk@itp.ac.cn}
	\affiliation{$^1$Institute of Theoretical Physics, Chinese Academy of Sciences, P.O. Box 2735, Beijing 100190, China}
	\affiliation{$^2$School of Physical Sciences, University of Chinese Academy of Sciences, No.19A Yuquan Road, Beijing 100049, China}
   \affiliation{$^3$Department of Physics, Yantai University, 30 Qingquan Road, Yantai 264005, China}
   \affiliation{$^4$School of Fundamental Physics and Mathematical Sciences, Hangzhou Institute for Advanced Study, University of Chinese Academy of Sciences, Hangzhou 310024, China}

\begin{abstract}
We study how environmentally driven spectral instabilities of quasinormal modes respond to parity violating gravito-scalar coupling in black holes.
Focusing on dynamical Chern-Simons gravity as a paradigm for parity violation, we perturb the Schwarzschild background with a localized potential bump. 
Our analysis reveals three distinctive phenomena absent in general relativity:
1) branch reconnections in the complex frequency plane,
2) a counterintuitive mode stabilization that delays overtaking transitions, and 
3) scalar mode dominance emerging at intermediate coupling strengths.
These frequency domain features show how comparatively weak static sector differences manifest as distinct dynamical signatures, thereby linking parity violating black hole perturbations with non-Hermitian spectral physics.
Our results provide a frequency domain characterization of parity violating coupling and motivate future targeted ringdown studies of modified gravity.
\end{abstract}

\maketitle

\section{Introduction}

Symmetries are cornerstones of general relativity (GR). 
In Schwarzschild spacetime, they provide a clean spectral baseline: axial and polar gravitational perturbations are isospectral and thus share identical quasinormal mode (QNM) spectra \cite{Regge:1957td,Chandrasekhar:1975nkd,Chandrasekhar:1985kt,Kokkotas:1999bd,Berti:2009kk}. 
However, many beyond GR theories, especially those with additional fields or parity selective couplings, can modify this structure \cite{Chen:2021cts,Li:2023ulk,del-Corral:2022kbk,Herceg:2024vwc,Herceg:2025fyf,Chung:2025gyg}.
Related questions also arise in theories that violate other fundamental symmetries, such as Lorentz symmetry \cite{Oliveira:2021abg,Zhang:2022fbz,Guo:2023nkd,Zhang:2025kcw}.
Determining whether such modifications leave identifiable imprints in black hole (BH) ringdown is a central goal of gravitational wave tests of strong field gravity \cite{Yagi:2017zhb,Perkins:2021mhb,Califano:2023aji,Zhu:2023rrx,Zhang:2025kcw}.
The challenge is that, on static BH backgrounds, such effects are often weak and may appear only as small spectral shifts or splittings, which are difficult to isolate observationally \cite{Chen:2021cts,del-Corral:2022kbk}.

A key insight is that BHs are not isolated Hermitian systems; 
they are open, dissipative systems whose QNM spectra are fundamentally non-Hermitian \cite{Kim:2005ts,Berti:2009kk,Arean:2023ejh,Besson:2025ghu}. 
Recent studies have shown that this non-Hermitian structure can make QNM spectra highly sensitive to small external perturbations, including generic environmental perturbations 
\cite{Leung:1997was,Arean:2023ejh,Cheung:2021bol,Solidoro:2024yxi,Besson:2025ghu,Shen:2025nsl,Cao:2025afs}. 
This sensitivity may manifest itself as spectral instabilities, in which QNM branches reorganize and the dominant mode changes discontinuously as the external perturbation is varied \cite{Cheung:2021bol}.

In this work, we use environmentally driven spectral sensitivity as a probe of parity violating gravito-scalar dynamics. 
As a concrete example, we consider dynamical Chern-Simons (dCS) gravity on a Schwarzschild background. 
Since the spherically symmetric background trivially decouples the polar sector from the scalar field \cite{Molina:2010fb,Cardoso:2009pk}, we can cleanly isolate the parity violating dynamics by tracking the axial gravito-scalar response against the polar benchmark.

To probe this response, we introduce a localized bump 
\footnote{
Here the localized bump is used as a perturbation of the effective gravitational potential. 
It parametrizes generic environmental or exterior structure perturbations that are known to reorganize QNM spectra, and is therefore useful for testing the spectral sensitivity of the axial gravito-scalar system.
The physical interpretation of the local bump can be found in Ref.\ \cite{Tian:2025uvk}.
}
in the effective gravitational potential as a perturbation \cite{Cheung:2021bol,Berti:2025hly} following the same spectral-instability setup used to study bump-induced QNM migration in Ref.\ \cite{Cheung:2021bol}, and track the migration of the QNM frequencies in the complex plane. 
The parameter $\beta$ controls the strength of the dCS induced coupling, with large $\beta$ corresponding to weak coupling.
Within this setup, we find three related phenomena: 
branch reconnections in the complex frequency plane, a postponement of the overtaking threshold for the dominant gravitational mode, and scalar dominance emerging at intermediate coupling strengths. 
These effects are absent in the polar perturbation and therefore inherently arise from the parity violating axial-scalar dynamics considered here.

These results place the problem in a broader non-Hermitian setting. 
In particular, the observed branch reconnections admit a natural reduced two-mode interpretation and are suggestive of exceptional point behavior in the coupled spectrum \cite{Motohashi:2024fwt,Cavalcante:2024swt,Kubota:2025hjk,Klaiman:2008zz,Schnabel:2017tti,Nesterov:2008cg,Kumar:2021lkq,Heiss:2012dx}.
At the same time, the present analysis should be viewed as a Schwarzschild benchmark study of frequency-domain spectral sensitivity. 
The quantities extracted below are spectral ones—branch migration, the switching threshold $a_{\rm crit}$, and the identity of the globally least-damped mode. 
As emphasized in \cite{Berti:2022xfj}, a pronounced rearrangement of the formal QNM spectrum need not imply an equally large prompt time-domain ringdown deformation. 
Our aim here is therefore to identify parity-sensitive spectral targets for future targeted ringdown spectroscopy and late-time mode extraction, rather than to provide a waveform forecast.
Extending the analysis to Kerr BHs, where dCS effects are astrophysically more relevant and odd parity gravito-scalar coupling persists, is an important next step \cite{Wagle:2021tam,Wagle:2023fwl,Li:2025fci}.

\section{Parity violating dynamical Chern-Simons model}

We utilize dCS gravity as a paradigmatic theory for gravity parity violation. 
The dCS action is given by \cite{Yunes:2009hc,Molina:2010fb,Boudet:2022wmb}
\begin{align}\label{eq:action}
	S_{\rm dCS}=  \int \dif^4 x \sqrt{-g}\ \bigg[& \frac{1}{2 \kappa} R +\frac{\alpha}{8} \phi(x) \varepsilon^{\mu \nu \rho \sigma} R_{\;\;\beta \mu \nu}^\alpha R_{\;\;\alpha \rho \sigma}^\beta \nonumber \\
    & -\frac{\beta}{2} \partial_\mu \phi \partial^\mu \phi\bigg],
\end{align}
Here we adopt geometric units where $c=G=1$. 
The background spacetime remains the standard Schwarzschild metric, 
$\dif s^2=-f(r)\dif t^2+f^{-1}(r)\dif r^2 + r^2 \dif {\mit \Omega}^2$ with $f(r)=1 - 2 M / r$,
and the Schwarzschild solution automatically corresponds to a vanishing background scalar field.
This choice is technically clean because the Pontryagin density vanishes on a spherically symmetric background, so Schwarzschild remains an exact background solution in dCS gravity \cite{Molina:2010fb,Cardoso:2009pk}.
However, the standard rotating Kerr solution cannot retain this property.
Following standard conventions, we set $\alpha=1$. 
The effective coupling strength is characterized by $1/\beta$:
large $\beta$ corresponds to weak axial-scalar coupling, while small $\beta$ signifies strong parity violating coupling.
Strictly speaking, the large-$\beta$ regime should be viewed as a weak coupling limit of the coupled system, in which the gravitational branch approaches the GR situation, while a scalar branch remains in the spectrum \cite{Molina:2010fb}.

For a vanishing background scalar field, the polar sector perturbations remain governed by the Zerilli equation, whereas the axial gravitational perturbation ($\mPsi$) is coupled to the dCS scalar perturbation ($\mTheta$) \cite{Molina:2010fb,Cardoso:2009pk}. 
The corresponding perturbation equations then take the coupled wave form \cite{Molina:2010fb}
\begin{widetext}
    \begin{subequations}\label{eq:pert-eq}
	\begin{equation}\label{eq:pert-eq-g}
		\frac{\dif^2 \mPsi}{\dif r_*^2}+\left[\omega^2-f\left(\frac{l (l +1)}{r^2}-\frac{6 M}{r^3}\right)\right] \mPsi=\frac{6 M}{r^5} f \mTheta,
	\end{equation}
	\begin{equation}\label{eq:pert-eq-s}
		\frac{\dif^2 \mTheta}{\dif r_*^2}+\left\{\omega^2-f\left[\frac{l (l +1)}{r^2}\left(1+\frac{36 M^2 }{r^6 \beta}\right)+\frac{2 M}{r^3}\right]\right\} \mTheta=f \frac{(l +2)!}{(l -2)!} \frac{6 M}{r^5 \beta} \mPsi,
	\end{equation}
\end{subequations}
\end{widetext}
where $r_*$ is the tortoise coordinate, defined by $\dif r_* \equiv \dif r/f(r)$.
Eq.\ \eqref{eq:pert-eq-s} shows explicitly that the scalar potential correction and the source term from $\mPsi$ are both controlled by $1/\beta$.
The functions $\mPsi$ and $\mTheta$ are derived from the Regge-Wheeler function and the scalar field expansion $\delta\phi=\dfrac{\mTheta(r)}{r}Y^{lm}(\theta,\varphi) \me^{-\mi\omega t}$, respectively.

It should be noted that in the $\beta \to \infty$ limit, while Eq.\ \eqref{eq:pert-eq-s} explicitly decouples into a free scalar field equation, the gravitational perturbation in Eq.\ \eqref{eq:pert-eq-g} retains an inhomogeneous source term driven by the scalar field. 
This implies that the eigenstates do not fully decouple, yet the quasinormal mode frequency spectrum of the system still asymptotically approaches the non-interacting union of the GR and pure scalar spectra. 
We will explicitly demonstrate the algebraic origin of this non-reciprocal limit using an upper triangular effective Hamiltonian matrix in the subsequent section.

We model the generic environmental perturbation by augmenting the standard effective potential with a localized P\"oschl-Teller barrier,
\begin{equation}
	V_{\rm eff}=V_{\rm Z/RW}+V_{\rm bump},\ V_{\rm bump}=\epsilon \sech^2\left(r_*-a\right).
\end{equation}
This bump simulates environmental effects or unknown physics \cite{Cardoso:2021wlq,Cardoso:2022whc}.
And $\epsilon$ measures the strength of the exterior perturbation, whereas $a$ specifies where this perturbation is placed relative to the main Regge-Wheeler/Zerilli barrier. 
Thus, for fixed $\epsilon$, varying $a$ scans the same exterior structure from the near-peak region to the farther tail, which is precisely the control procedure used to expose bump-induced spectral instability in Ref.\ \cite{Cheung:2021bol}.
It is introduced independently of the dCS coupling and is not meant to represent a localized parity violating region outside the BH.
In our numerical calculations of the next section, we fix $M=1$, $l=2$, and $\epsilon=10^{-2}$. 
We then systematically track the QNM frequencies in the complex plane as the position of bump $a$ is varied, moving away from the peak of the unperturbed Regge-Wheeler/Zerilli barrier ($r_{*} \simeq 1.614$), which serves merely as a spatial reference point.
Our choice of amplitude $\epsilon=10^{-2}$ is a deliberate strategy to ensure numerical tractability. 
As established in Ref.\ \cite{Cheung:2021bol}, the critical position $a_{\rm crit}$ for instability is inversely related to $\epsilon$, cf.\ Fig.\ 2 of that work.
A physically plausible, smaller $\epsilon$ would simply push the same critical phenomena to numerically challenging large distances. 
We therefore use $\epsilon=10^{-2}$ to bring the relevant dynamics into a numerically reliable computational domain, without changing the physical mechanism under investigation.

\section{Spectral Dynamics}

Building on the model introduced in the previous section, we now present the main numerical results for the QNM spectrum of the polar sector and of the axial gravito-scalar system in the presence of the bump perturbation. 
We employ the shooting method to solve the corresponding QNM eigenfrequencies $\omega_n$ of the polar Zerilli equation and the axial coupled equations Eq.\ \eqref{eq:pert-eq} after the inclusion of the bump potential.
Specifically, this method imposes pure ingoing wave conditions at the event horizon and pure outgoing waves at spatial infinity, with the latter enforced via a numerical cutoff. 
The complex frequencies are found using Mathematica's \texttt{FindRoot} function. 
We confirmed that all key phenomena discussed below are obtained with residuals of $10^{-6}$ or smaller.

\begin{figure*}[!htb]
	\includegraphics[width=0.7\linewidth]{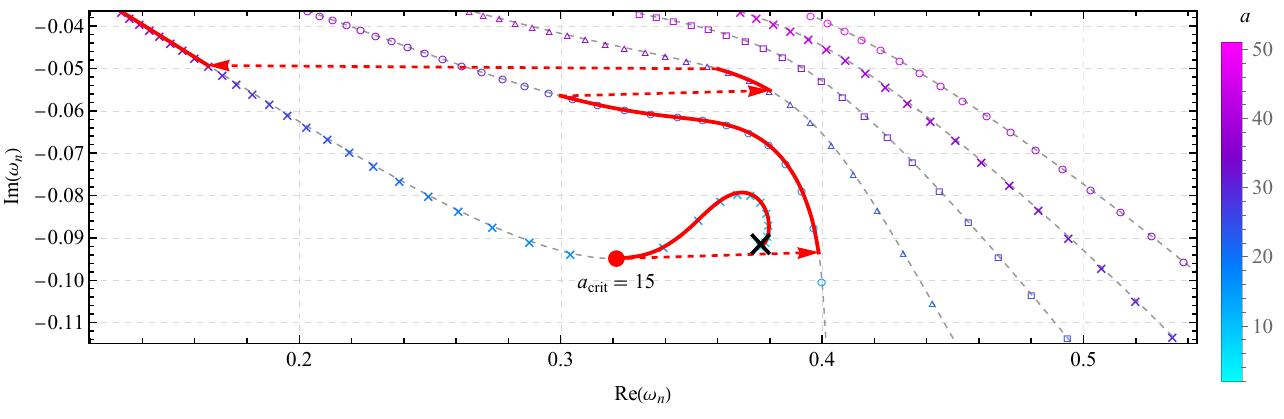}
	\caption{
	QNM migration trajectories in the complex frequency plane for the polar benchmark under the same bump perturbation.
	The discrete points are the numerically computed QNM frequencies, dashed segments connect the same branch as $a$ is varied, and the marker shapes ($\times$, $\circ$, $\triangle$, $\square$) distinguish different branches. 
    Specifically, the black crosses mark the starting points of the migrations at $a=2$.
    At this starting point the bump lies close to the photon-sphere potential peak and effectively merges with the main barrier, so its influence on the QNM spectrum is small and the frequencies are near the Schwarzschild convergence point of the scan; we therefore use $a=2$ as the starting point of the scan.
    The color bar gives the value of $a$ along each branch. 
    The red curve marks the dominant mode $\varpi(a)$ defined in the text, and the arrows mark the jumps of $\varpi(a)$ as $a$ is increased; the label $a_{\rm crit}$ marks the first branch switch of this dominant track.
    The dominance comparison is made at fixed $a$, by comparing all resolved branches carrying the same value of $a$.
	}
	\label{fig:polar-migration}
\end{figure*}

\begin{figure*}[!htb]
    \includegraphics[width=\linewidth]{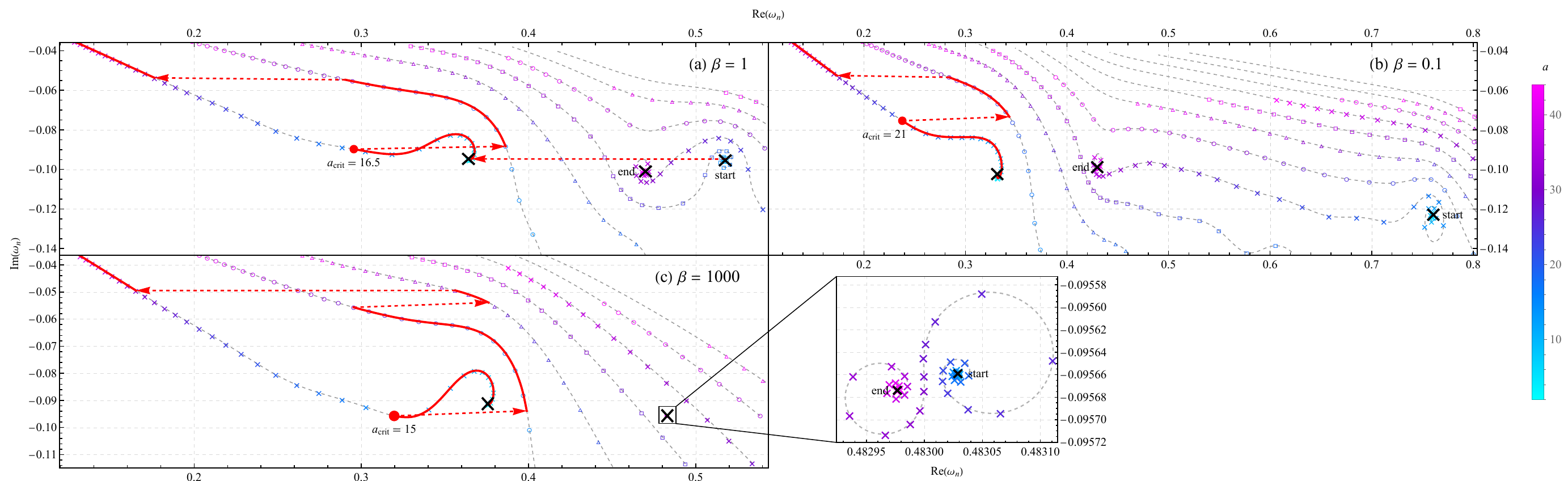}
    \caption{
    QNM migration trajectories in the complex frequency plane for different coupling strengths $\beta$. 
	The same plotting conventions as in Fig.\ \ref{fig:polar-migration} are used here. 
	In particular, the red track in each panel should be read using the same fixed-$a$ dominance criterion as in the polar benchmark.
	Within each panel, the spectrum is generated by varying $a$ at fixed $\beta$. 
    The black crosses label the two ends of the $a$ scan: the ``start'' cross ($a=2$) corresponds to the bump merging with the main barrier, while the ``end'' cross marks the termination of the scan at large $a$, toward which the spectral lines eventually converge.
    As in the polar case, the start cross is a near-Schwarzschild reference point where the bump is close to the photon-sphere barrier and produces only a small QNM shift.
	Across panels, one compares how the whole set of branch trajectories is reshaped when the axial-scalar coupling changes.
    For large $\beta$, the gravitational trajectories approach the polar benchmark, whereas at intermediate and strong coupling the branch connectivity is reorganized.}
	\label{fig:migration}
\end{figure*}

\begin{figure}[!htb]
	\includegraphics[width=\linewidth]{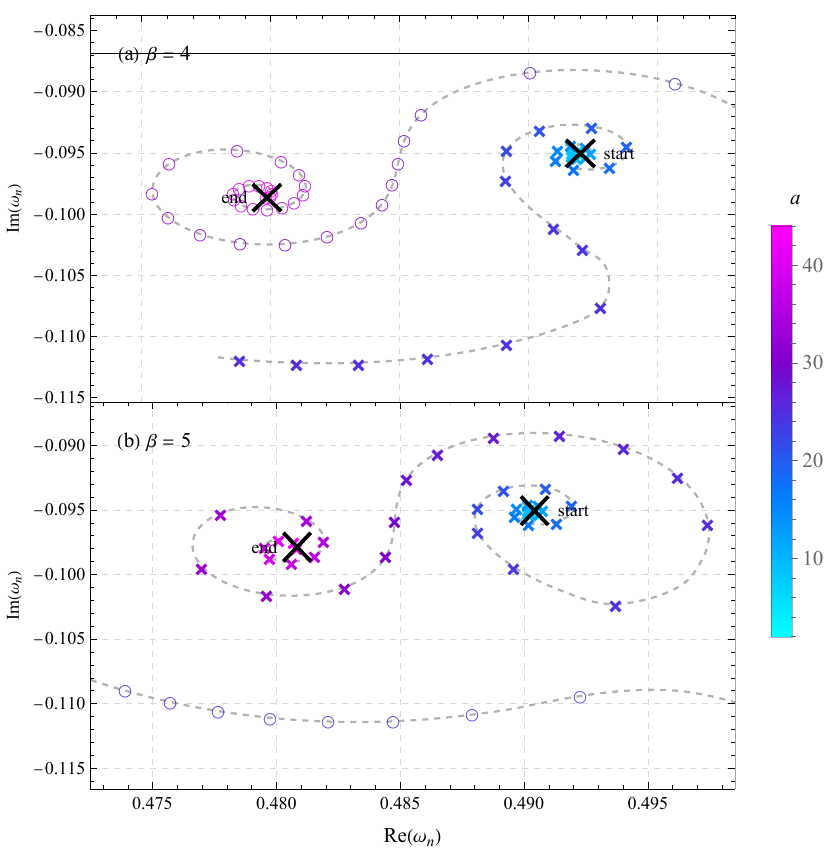}
	\caption{Fine scan of the relevant migration branches near the reconnection region for $\beta = 4$ and $5$. 
    The connectivity of the two curves is exchanged across this interval, indicating a branch reconnection event.}
	\label{fig:merge-1}
\end{figure}

To explore these effects, we compute the spectra for $\beta = 0.1,\ 1$, and $1000$, and compare them with the polar sector, which serves here as a benchmark.
The corresponding migration trajectories of this benchmark are shown separately in Fig.\ \ref{fig:polar-migration}.
Throughout Figs.\ \ref{fig:polar-migration}-\ref{fig:merge-1}, we keep $\epsilon=10^{-2}$ fixed and vary only the bump position $a$ within each panel. 
Each family of discrete points connected by a dashed curve is therefore the migration trajectory of one QNM branch as $a$ changes at fixed $\beta$; 
comparing different panels means comparing different spectra at different $\beta$, not the continuation of a single mode across $\beta$. 
The marker shapes ($\times$, $\circ$, $\triangle$, $\square$) distinguish different branches, and the color attached to each point records the value of $a$ along that branch. 
Specifically, the black crosses mark the starting points of the migrations at $a=2$.
At this location, the bump is situated near the peak of the effective potential, effectively merging with the main barrier rather than forming a distinct secondary cavity; 
consequently, the QNM frequencies almost recover the unperturbed vacuum values \cite{Cheung:2021bol}.
At each fixed $a$, we denote by $\varpi(a) \equiv \arg \min_n |\Im \omega_n(a)| $ the globally dominant mode, i.e. the stable QNM closest to the real axis; equivalently, since stable modes have $\Im\omega_n<0$, it is the mode with the largest $\Im\omega_n$. 
This minimization is performed at fixed $a$: the modes to be compared are all QNMs of the same static potential $V_{\rm eff}(r_*;a)$, not different points along a single migration curve.
Since $a$ is encoded by color rather than by either axis of the complex-frequency plot, modes with the same value of $a$ need not lie on a horizontal or vertical line in the figure.
The red track in the migration plots is obtained by following this dominant mode through the full spectrum as $a$ varies. 
A discontinuous jump of this red track indicates that the dominant mode has switched branch identity, and $a_{\rm crit}$ is the first value of $a$ at which this happens. 
Throughout the paper, this ``instability'' refers to this spectral switching instability of the dominant branch, not to a dynamical instability with $\Im(\omega)>0$.

Further analysis of axial sector will systematically trace how the quasinormal frequencies $\omega_n$ respond to the bump position $a$ and the parameter $\beta$ in the complex frequency space. 
Our focus is on three observables, the migration of the QNM branches, the critical position $a_{\rm crit}$ at which the dominant gravitational branch first switches, and the frequency deviation of the post overtaking dominant mode relative to the vacuum value,
$\Delta\omega \equiv \omega(\epsilon,a)-\omega(0)$.

For weak coupling $\beta = 1000$, the trajectories closely mirror those in GR.
In particular, the gravitational migration curves in Fig.\ \ref{fig:migration}(c) are numerically almost indistinguishable from the polar benchmark trajectories shown in Fig.\ \ref{fig:polar-migration}.
This is consistent with the weak coupling structure of the coupled system: 
at large $\beta$, the induced scalar response along the gravitational branch is strongly suppressed, so that the corresponding axial trajectories approach the polar benchmark.
At the same time, the scalar branches remain present in the full spectrum, but their coupling with the gravitational branches is weak.
Specifically in the complex frequency space, the migration path of each scalar mode starts from a vortex and quickly terminates in another vortex. 
They are completely separated from the migration paths of the gravitational modes, as shown in the panel(c) of Fig.\ \ref{fig:migration}.

As the coupling strengthens, the migration pattern departs markedly from this weak coupling behavior. 
We then observe a series of branch reconnections, in which the connectivity of neighboring migration curves changes discontinuously as $\beta$ is varied.
This is first seen by comparing the macro-scale spectrum at $\beta=1000$ (Panel (c) of Fig.\ \ref{fig:migration}) with the spectrum at $\beta=1$ (Panel (a) of Fig.\ \ref{fig:migration}).
At high $\beta$, we focus on two distinct, separate curves: an S-shaped curve containing both vortices corresponds to the dCS scalar mode and a simple curve corresponds to a GR overtone.
At $\beta=1$, however, the morphology and relative positions of these two curves have significantly changed. 
To locate this reconnection event, we performed a fine scan of parameter $\beta$, detailed in Fig.\ \ref{fig:merge-1}. 
At $\beta=5$ (Panel(b) of Fig.\ \ref{fig:merge-1}), the spectrum consists of at least two distinct curves: a simple curve (henceforth curve A) and an S-shaped composite curve containing two vortices (henceforth curve B). 
As $\beta$ crosses the critical value and reaches $\beta=4$ (Panel(a) of Fig.\ \ref{fig:merge-1}), a reconnection occurs: the first part of curve A now connects to the second part of curve B, while the first part of curve B connects to the second part of curve A.
This abrupt exchange of connectivity is the branch reconnection phenomenon responsible for the macro-scale rearrangement seen in Fig.\ \ref{fig:migration}.

To make the above reconnection picture more explicit, we now introduce a reduced non-Hermitian description aimed at elucidating the mechanism behind the branch reconnection phenomenon.
Since the critical overlap integrals depend non-trivially on both $\beta$ and $a$, this construction is intended as a qualitative explanation of exceptional point behavior rather than a quantitative predictor of their exact locations.

Technically, the perturbed state vector $\Phi$ evolves in the space of perturbations spanned by the relevant unperturbed GR and scalar modes.
However, the spectral instabilities of interest—such as curve merging and mode reconnection—arise from the strong resonant interaction between specific mode pairs.
Justified by this spectral selectivity, we adopt the two-mode approximation, neglecting the off-resonant couplings to distant spectator modes and projecting the full operator onto the relevant subspace spanned by a specific gravitational mode $|\psi_g^{(n)}\rangle$ and a scalar mode $|\psi_s^{(k)}\rangle$.

Accordingly, we apply a reduced non-Hermitian perturbative analysis \cite{Heiss:2012dx,Yang:2025dbn} for the coupled master equations
\begin{equation}\label{eq:eig-eq}
    \hat{\mathcal{L}}(\beta, a) \Phi = \omega^2 \Phi.
\end{equation}
By rewriting the coupled wave equations Eq.\ \eqref{eq:pert-eq} into the matrix form, we identify the specific forms of these operators acting on the state vector $\Phi=(\mPsi, \mTheta)^{\rm T}$.
The diagonal unperturbed operators correspond to the standard Regge-Wheeler and scalar potentials in Schwarzschild spacetime
\begin{subequations}
    \begin{equation}
    \hat{H}_g^{\rm GR} \equiv -\frac{\dif^2}{\dif r_*^2} + f \left(\frac{l(l+1)}{r^2} - \frac{6M}{r^3}\right),
    \end{equation}
    \begin{equation}
        \hat{H}_s^{\rm GR} \equiv -\frac{\dif^2}{\dif r_*^2} + f \left(\frac{l(l+1)}{r^2} + \frac{2 M}{r^3}\right).
    \end{equation}
\end{subequations}
The off-diagonal operators capture the parity violating coupling.
Comparison with Eq.\ \eqref{eq:pert-eq-g} identifies the gravitational-to-scalar coupling
\begin{equation}
    \hat{\mathcal{C}}_{gs} \equiv \frac{6 M f}{r^5}.
\end{equation}
Conversely, Eq.\ \eqref{eq:pert-eq-s} reveals that both the scalar-to-gravitational coupling and the dCS correction to the scalar potential carry a $\beta^{-1}$ scaling, allowing us to factor out the parity violating parameter
\begin{equation}
    \hat{\mathcal{C}}_{sg} \equiv f \frac{(l+2)!}{(l-2)!} \frac{6M}{r^5}, \quad
    \hat{\mathcal{P}}_{\rm dCS} \equiv f \frac{l(l+1)}{r^2} \frac{36M^2}{r^6}.
\end{equation}
Substituting these operators into Eq.\ \eqref{eq:eig-eq}, we decompose the operator into a diagonal unperturbed part, a parity violating interaction $\hat{V}_{\rm dCS}(\beta)\equiv \hat{\mathcal{P}}_{\rm dCS}/\beta$, and an environmental perturbation $\hat{V}_{\rm bump}(a)$
\begin{equation}
    \hat{\mathcal{L}} =
    \begin{pmatrix}
        \hat{H}_g^{\rm GR} & 0 \\
        0 & \hat{H}_s^{\rm GR}
    \end{pmatrix}
    +
    \begin{pmatrix}
        \hat{V}_{\rm bump}(a) & \hat{\mathcal{C}}_{gs} \\
        \beta^{-1}\hat{\mathcal{C}}_{sg} & \beta^{-1}\hat{\mathcal{P}}_{\rm dCS}
    \end{pmatrix}.
\end{equation}
It is evident from this matrix representation that in the $\beta \to \infty$ limit, the operator $\hat{\mathcal{L}}$ strictly reduces to an upper triangular matrix rather than a purely diagonal one. 
However, since the eigenvalues of an upper triangular matrix are determined entirely by its diagonal elements, the quasinormal mode eigenfrequency spectrum of the system is algebraically equivalent to the non-interacting union of the GR axial gravitational spectrum and the pure scalar spectrum ($\{\omega_g^{\rm GR}\} \cup \{\omega_s^{\rm GR}\}$). 
This algebraic property shows that, in the large $\beta$ weak coupling limit, the gravitational subset of the spectrum approaches the GR benchmark, while the full coupled spectrum tends to the union of the GR gravitational and free scalar spectra.
We include this remark only to clarify the weak-coupling spectral limit of the coupled equations; it is not used below as an independent physical claim.

By projecting the operator onto the subspace of the unperturbed $n$-th gravitational mode $|\psi_g^{(n)}\rangle$ and $k$-th scalar mode $|\psi_s^{(k)}\rangle$, we derive the local effective Hamiltonian
\begin{equation}
    H_{\rm eff} =
    \begin{pmatrix}
        \Omega_g^2(a) & \kappa_{gs} \\
        \kappa_{sg}(\beta) & \Omega_s^2(\beta)
    \end{pmatrix}.
\end{equation}
Here, the diagonal elements capture the bare frequencies renormalized by their respective perturbations.
The gravitational eigenvalue explicitly depends on the bump position via the overlap integral
\begin{equation}
    \Omega_g^2(a) \equiv \omega_{g, {\rm GR}}^{(n)2} + \langle \psi_g^{(n)} | \hat{V}_{\rm bump}(a) | \psi_g^{(n)} \rangle,
\end{equation}
where the bracket denotes a renormalized unconjugated bilinear form appropriate for open BH QNM systems.
Its explicit construction is subtle because QNM wavefunctions are not square-integrable under the physical boundary conditions, and in general requires a regularization prescription; see Refs.\ \cite{Leung:1997was,Green:2022htq} for representative constructions.
Here we use this bracket only as a compact notation for the reduced mode projection.
The scalar eigenvalue describes the continuous spectral shift driven by the dCS potential
\begin{equation}
    \Omega_s^2(\beta) \equiv \omega_{s, {\rm GR}}^{(k)2} + \frac{1}{\beta} \langle \psi_s^{(k)} | \hat{\mathcal{P}}_{\rm dCS} | \psi_s^{(k)} \rangle.
\end{equation}
The off-diagonal terms represent the effective non-Hermitian couplings
\begin{equation}
    \kappa_{gs} \equiv \langle \psi_g^{(n)} | \hat{\mathcal{C}}_{gs} | \psi_s^{(k)} \rangle, \quad
    \kappa_{sg}(\beta) \equiv \frac{1}{\beta}\langle \psi_s^{(k)} | \hat{\mathcal{C}}_{sg} | \psi_g^{(n)} \rangle.
\end{equation}

Within the reduced description, the discriminant $\mathcal{D}$ of the characteristic polynomial $\det(H_{\rm eff} - \lambda \mathbf{I}) = 0$ provides a useful local diagnostic.
The condition $\mathcal{D}=0$ identifies an approximate exceptional point locus in the parameter space
\begin{equation}\label{eq:resonance-cond}
    \mathcal{D}(\beta, a) =
    \left[ \Omega_g^2(a) - \Omega_s^2(\beta) \right]^2
    + 4 \kappa_{gs}\kappa_{sg}(\beta) = 0.
\end{equation}
This algebraic constraint suggests that for a specific mode pair $(n,\ k)$, the reconnection region is confined to a corresponding trajectory in the $(\beta, a)$ parameter space.
It requires that the squared frequency detuning, $\Omega_g^2(a) - \Omega_s^2(\beta)$, be balanced by the non-Hermitian effective coupling product.
In this sense, varying the control parameter $\beta$ can bring the scalar sector into critical resonance with the gravitational spectrum and thereby provide a qualitative explanation of the branch reconnections seen in the full numerical results.
The observed reconnection is therefore consistent with critical behavior in a reduced non-Hermitian description \cite{Cavalcante:2024swt} near exceptional point, without implying that the full spectrum is globally controlled by a single $2 \times 2$ model.
A similar event is also suggested by comparing panel(a) and (b) of Fig.\ \ref{fig:migration}.

A second quantitative discovery comes from the critical position for mode overtaking, $a_{\rm crit}$—the value of $a$ at which the first discontinuous jump in the dominant mode $\varpi$ occurs. 
This is the same spectral-instability threshold introduced in Ref.\ \cite{Cheung:2021bol}, now applied to the coupled axial-scalar system.
We treat the polar sector as the GR baseline; according to Fig.\ \ref{fig:polar-migration}, it yields an overtaking threshold of $a_{\rm crit}^{\rm polar} \simeq 15$.
For the axial gravito-scalar system, the threshold depends clearly on $\beta$:
$\beta=1000$ gives $a_{\rm crit}\simeq 15$, numerically coincident with the polar baseline;
$\beta=1$ shifts the threshold to $a_{\rm crit}\simeq 16.5$;
and $\beta=0.1$ further postpones it to $a_{\rm crit}\simeq 21$.
Thus, stronger coupling (smaller $\beta$) significantly delays the switching instability, with $a_{\rm crit}$ increasing from $\simeq 15$ to $\simeq 21$ at $\beta=0.1$.
Because the scan parameter is the bump position, this means that a stronger dCS coupling requires the same exterior structure to be placed farther from the main barrier before the globally dominant branch is forced to switch.
This counterintuitive dynamical stabilization arises because the dCS scalar potential $\propto 1/(\beta r^6)$
becomes steeply repulsive at strong coupling, spatially expelling the scalar perturbation $\mTheta(r)$ to larger radii.
As a consequence, the near zone overlap between the gravitational and scalar states is reduced, and so is the effective hybridization that drives the switching instability.
In the language of the reduced model, the relevant overlap integrals entering $\kappa_{sg}(\beta)$ are weakened until the bump is moved farther into the exterior region.
This points to a non-perturbative mechanism: the parity violating coupling does not generically destabilize the dominant gravitational branch.
Instead, a coupling dependent suppression of near field hybridization delays its switching instability \cite{Bojowald:2024lhr}.

\begin{figure}[!htb]
	\includegraphics[width=\linewidth]{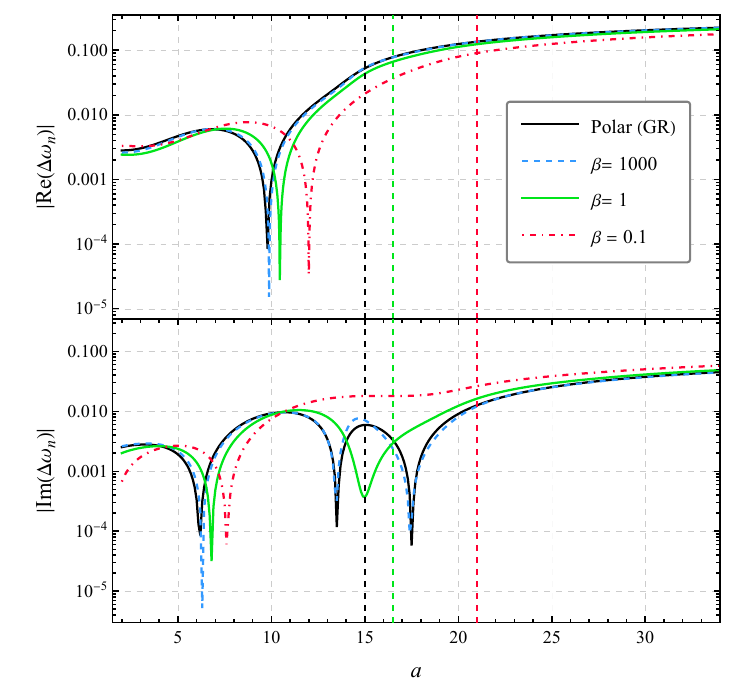}
	\caption{The frequency deviation of the dominant mode $\Delta\omega=\omega(\epsilon,a)-\omega(0)$ as a function of the bump position $a$, focusing on the post overtaking stable regime ($a\gtrsim a_{\rm crit}$). The plotted quantities are absolute values and therefore emphasize the magnitude of the frequency and damping shifts.
    }
	\label{fig:delta_omega}
\end{figure}

Fig.\ \ref{fig:delta_omega} depicts the real and imaginary parts of $\Delta \omega$ as functions of $a$. We focus on the post overtaking stable regime with large $a$.
It is proposed to compare the size of the spectral response for different $\beta$. 
Since $\epsilon$ is fixed throughout, the horizontal axis $a$ is again the environmental control variable: 
it tells how the same exterior perturbation changes the dominant frequency as the bump is moved from the near-peak region to the outer tail, while $\beta$ determines how strongly the axial gravito-scalar system reacts to that displacement.
For the real part $\left|\Re(\Delta\omega)\right|$, the polar sector shows the largest deviation. 
As parity violation strengthens, the offset is progressively suppressed, indicating the scalar coupling mitigates frequency shifts.
Conversely, for the imaginary part $\left|\Im(\Delta\omega)\right|$, an inversion pattern is observed, weak coupling exhibits the smallest offset, and it is close to the GR situation. 
As coupling strengthens, the damping offset significantly increases, with strong parity violation showing the largest deviation. 
It reveals a twofold effect of parity violation,
\begin{enumerate}
	\item On overtaking, strong parity violation stabilizes the system, postponing $a_{\rm crit}$;
	\item On asymptotic behavior, strong parity violation makes the resulting stable mode more sensitive to perturbations in the potential's tail.
\end{enumerate}

In qualitative terms, stronger coupling makes the real part of the spectrum more rigid against distant perturbations, while simultaneously increasing the susceptibility of the damping rate to tail modifications \cite{Garcia-Saenz:2024beb}.
This mechanism, where energy is absorbed by the scalar field to alleviate damping, parallels the stabilization of overtones by scalar hair in Einstein-scalar-Gauss-Bonnet models \cite{Antoniou:2017hxj}. 
Relative to the polar benchmark, this asymmetric $\beta$-dependence provides a compact dynamical characterization of the parity violating coupling.

Another entirely new phenomenon, scalar dominance, arises for $\beta = 1$ at small $a \lesssim 5$. 
Most strikingly, at intermediate coupling ($\beta \simeq 1$) and near peak bumps ($a \lesssim 5$),
the dCS pseudoscalar mode becomes the globally longest-lived mode in the entire spectrum, a behavior absent in the polar benchmark and weak coupling regime (see red segments in panel(a) of Fig.\ \ref{fig:migration} where the dominant mode track $\varpi$ abruptly switches to the scalar branch for $a \lesssim 5$). 
This suggests that the late time signal may transiently align with a scalar branch in this parameter window.
Whether such a spectral rearrangement leaves a clean time domain observable, however, requires a dedicated waveform analysis and cannot be inferred from the present frequency domain results alone.

In the intermediate coupling regime, the scalar dominance can be understood within the same reduced non-Hermitian Hamiltonian introduced above.
To quantify this effect, we now apply second-order perturbation theory to $H_{\rm eff}$ in the non-degenerate regime, where
\begin{equation}
    \left|\Omega_s^2(\beta)-\Omega_g^2(a)\right| \gg |\kappa_{gs}|,\ |\kappa_{sg}(\beta)|.
\end{equation}
We decompose the effective Hamiltonian as
\begin{subequations}
    \begin{equation}
    H_{\rm eff}=H_0+V,
\end{equation}
where
\begin{equation}
    H_0=
    \begin{pmatrix}
        \Omega_g^2(a) & 0\\
        0 & \Omega_s^2(\beta)
    \end{pmatrix},
    \quad
    V=
    \begin{pmatrix}
        0 & \kappa_{gs}\\
        \kappa_{sg}(\beta) & 0
    \end{pmatrix}.
\end{equation}
\end{subequations}
The zeroth-order scalar eigenvalue is therefore
\begin{equation}
    \lambda_s^{(0)}=\Omega_s^2(\beta),
\end{equation}
with right eigenvector $|s^{(0)}\rangle=(0,1)^{\rm T}$.
In this reduced basis, the corresponding left eigenvector is the same canonical basis vector, so the standard non-Hermitian perturbation formula applies directly.
Because the perturbation $V$ is purely off-diagonal, the first-order correction vanishes identically,
\begin{equation}
    \Delta\lambda_s^{(1)}=
    \langle s^{(0)}|V|s^{(0)}\rangle=0.
\end{equation}
The leading correction thus appears at second order and is given by
\begin{equation}
    \Delta\lambda_s^{(2)} = \frac{H_{21}H_{12}}{H_{22}-H_{11}}
    =
    \frac{\kappa_{sg}(\beta)\kappa_{gs}}
         {\Omega_s^2(\beta)-\Omega_g^2(a)},
\end{equation}
where $\lambda_g^{(0)}=\Omega_g^2(a)$ is the zeroth-order gravitational eigenvalue.
Equivalently, solving the characteristic equation of the $2\times 2$ matrix and expanding around the scalar branch yields the same result.
The scalar complex eigenvalue $\lambda_s=\omega_s^2$ is therefore
\begin{equation}
    \lambda_s
    \simeq
    \Omega_s^2(\beta)
    +
    \frac{\kappa_{sg}(\beta)\kappa_{gs}}
         {\Omega_s^2(\beta)-\Omega_g^2(a)}.
\end{equation}
To convert this result from $\lambda_s=\omega_s^2$ to the frequency itself, we expand $\omega_s=\sqrt{\lambda_s}$ around the unperturbed scalar frequency $\Omega_s(\beta)$.
For a small shift $\Delta\lambda_s\equiv \lambda_s-\Omega_s^2(\beta)$, one has
\begin{equation}
    \omega_s
    =
    \sqrt{\Omega_s^2(\beta)+\Delta\lambda_s}
    \simeq
    \Omega_s(\beta)+\frac{\Delta\lambda_s}{2\Omega_s(\beta)}.
\end{equation}
Substituting the second-order result then gives
\begin{equation}
    \omega_s
    \simeq
    \Omega_s(\beta)
    +
    \frac{1}{2\Omega_s(\beta)}
    \frac{\kappa_{sg}(\beta)\kappa_{gs}}
         {\Omega_s^2(\beta)-\Omega_g^2(a)}.
\end{equation}
Hence the decay rate is approximately
\begin{equation}\label{eq:decay-rate-rigorous}
    \Im(\omega_s)
    \simeq
    \Im\left[\Omega_s(\beta)\right]
    +
    \Im \left[
        \frac{1}{2\Omega_s(\beta)}
        \frac{\kappa_{sg}(\beta)\kappa_{gs}}
             {\Omega_s^2(\beta)-\Omega_g^2(a)}
    \right].
\end{equation}
Assuming the static dCS potential correction predominantly shifts the real part of the frequency, we can approximate $\Im[\Omega_s(\beta)] \simeq \Im(\omega_s^{\rm GR})$ and use the simpler estimate
\begin{equation}\label{eq:decay-rate}
    \Im(\omega_s)
    \simeq
    \Im(\omega_s^{\rm GR})
    +
    \Im\!\left[
        \frac{\kappa_{sg}(\beta)\kappa_{gs}}
             {\Omega_s^2(\beta)-\Omega_g^2(a)}
    \right].
\end{equation}
This perturbative result clarifies the strong spatial selectivity of the scalar dominance.
The effect is strongest for near field bumps (small $a$), because the overlap integrals $\kappa_{gs}$ and $\kappa_{sg}$ are dominated by the short range coupling terms $\sim r^{-5}$.
At the same time, the effect is suppressed again at very strong coupling, such as $\beta=0.1$, because the stiffened dCS scalar barrier pushes the scalar wavefunction outward and thereby reduces the effective overlap with the gravitational sector despite the explicit $\beta^{-1}$ enhancement.

This scalar dominance, absent in GR or weak coupling cases, represents a parity violation induced mode crossing. 
Here, the dCS pseudoscalar field $\phi$ invades and stabilizes the axial dynamics near the potential peak. 
The scalar mode gains enhanced longevity under near peak bumps because of the strong chiral coupling. 
The absence of this overtake in the polar perturbation highlights the sector selectivity of the dCS coupling.
It therefore identifies a qualitatively distinct spectral signature of parity violating gravito-scalar dynamics.

\section{Conclusion and Discussion}

In this work, we have studied how parity violating gravito-scalar coupling in dCS gravity modifies the spectral response of a Schwarzschild BH to an external bump perturbation.
Using the polar sector as a GR benchmark, we identified three qualitative features of the coupled axial-scalar dynamics: 
branch reconnections in the complex frequency plane, a counterintuitive stabilization of the dominant gravitational branch through the increase of the critical switching threshold $a_{\rm crit}$ at stronger coupling, and scalar dominance near the potential peak at intermediate coupling.
Within the present Schwarzschild benchmark, these features appear only in the axial gravito-scalar system; 
among them, the delayed switching threshold and scalar dominance are more directly attributable to the parity violating coupling, while the branch reconnections admit a broader non-Hermitian interpretation.
These findings show that comparatively weak static sector differences can be reorganized into substantial branch level effects in the frequency domain.
Within the setup considered here, strong coupling, e.g. $\beta \sim 0.1$, shifts the switching threshold to a stabilization window roughly given by $15 \lesssim a \lesssim 21$, where the dominant gravitational branch remains stable beyond the GR benchmark.

From the observational viewpoint, we focus on spectral dynamics rather than full waveform forecasts. 
In our setup, the localized bump acts as a scattering barrier that sources GW echoes.
Through dispersive superposition over multiple round trips, these echoes eventually evolve into steady-state cavity resonances \cite{Mark:2017dnq,Maggio:2019zyv,Hu:2025beh}. 
Therefore, the primary imprint of parity violation is not a manifest deformation in the prompt ringdown, but rather a modified late-time spectral pattern—most notably, a delayed switching threshold and a temporary scalar-led branch. 
Interpreting the least-damped spectral branch as the dominant late-time waveform component assumes that the relevant excitation residues are of comparable order and are not parametrically suppressed.
Fitting these late-time cavity modes are the precise target quantities for future high-fidelity spectroscopy analyses. 
This physical picture aligns with recent time-domain studies \cite{Berti:2022xfj,Spieksma:2024voy,Hu:2026sxj}, demonstrating that strong spectral instability does not necessarily imply a comparably large modification to the prompt waveform.
Accordingly, the present frequency domain results alone do not establish a direct observational degeneracy or separability at the waveform level.
We therefore refrain from making a direct detector level claim here; instead, the shift of $a_{\rm crit}$ and the scalar dominance should be viewed as concrete target features for future time domain and data analysis studies.
In this sense, a delayed switching threshold relative to the polar benchmark provides a clean frequency domain characterization of the parity violating coupling in the present model.
Crucially, as detailed in the App.\ \ref{app:A}, the critical point follows an approximately logarithmic scaling with the perturbation amplitude, $a_{\rm crit} \propto -\log \epsilon$.
This suggests that the same stabilization mechanism persists for weaker external perturbations, while the corresponding critical window is displaced to larger radii.

A natural and imperative progression of this work is its application to spinning Kerr BHs.
This extension is particularly important because astrophysical merger remnants are spinning, and rotation itself can act as a non-Hermitian control parameter, lifting the $m$-mode degeneracy and potentially driving the spectrum toward exceptional point behavior \cite{Cavalcante:2024swt}.
We therefore expect a nontrivial interplay between the dCS induced axial-scalar coupling and the BH spin $a/M$, especially since slowly rotating and more precise spinning dCS QNM spectra have now become available \cite{Wagle:2021tam,Li:2025fci,Chung:2025gyg}.
Future work should also distinguish the physical origins of these phenomena; while branch reconnections naturally arise from generic non-Hermitian two-field couplings \cite{Takahashi:2025uwo}, the counterintuitive stabilization and scalar dominance appear to be specific dynamical signatures attributable to the parity violating mechanism.

From a phenomenological perspective, the present results do not by themselves place new bounds on the dCS coupling. 
Rather, they isolate frequency-domain mechanisms that may inform future targeted ringdown studies of dCS gravity and complement existing constraints \cite{Silva:2020acr,Perkins:2021mhb,Wagle:2021tam}.
More broadly, the present work demonstrates that within the framework of parity-violating dCS gravity, non-Hermitian spectral sensitivity induces a distinct structural transition in the QNM spectrum under environmental perturbations. 
This systematic response provides a concrete phenomenological signature to separate parity-violating couplings from standard general-relativistic environmental effects, while the exact translation from spectral structure to observable ringdown signatures must be established case by case.

\section*{Acknowledgments}

We thank the anonymous referees for their valuable comments and suggestions.
This work is supported by the National Natural Science Foundation of China No.~12475067 and No.~12235019. Moreover, C.Lan is supported by Yantai University under Grant No.~WL22B224.

\appendix
\section{Scaling law of $a_{\rm crit}$ with perturbation amplitude $\epsilon$}\label{app:A}

To validate the generality of the perturbation amplitude $\epsilon = 10^{-2}$ used in the main text for numerical tractability, and to exclude the possibility that the observed critical phenomenon is an artifact of this large $\epsilon$ value, we here investigate the scaling relation of $a_{\rm crit}$ with $\epsilon$. 
We track $a_{\rm crit}$ for the dCS axial mode with $\beta=1$ as $\epsilon$ is varied from $10^{-2}$ down to $10^{-5}$.

Our numerical results for the $\beta=1$ case, as shown in Tab.\ \ref{tab:scaling}, clearly demonstrate a robust trend: as the perturbation amplitude $\epsilon$ decreases, the critical position $a_{\rm crit}$ is systematically pushed to larger distances. 
Despite minor numerical fluctuations at intermediate data points, the data exhibits a strong linear relationship on a semilogarithmic plot.

\begin{table}[!htb]
\centering
\caption{Scaling relation data for the critical position $a_{\rm crit}$ as a function of the bump amplitude $\epsilon$.}
\label{tab:scaling}
\begin{tabular}{lccccccc}
\toprule
$\epsilon$ & $10^{-2}$ & $10^{-2.5}$ & $10^{-3}$ & $10^{-3.5}$ & $10^{-4}$ & $10^{-4.5}$ & $10^{-5}$ \\
\midrule
$a_{\rm crit}$ & 16.5 & 27 & 25 & 34.5 & 33.5 & 43 & 42.5 \\
\bottomrule
\end{tabular}
\end{table}

\begin{figure}[!htb]
    \centering
    \includegraphics[width=\linewidth]{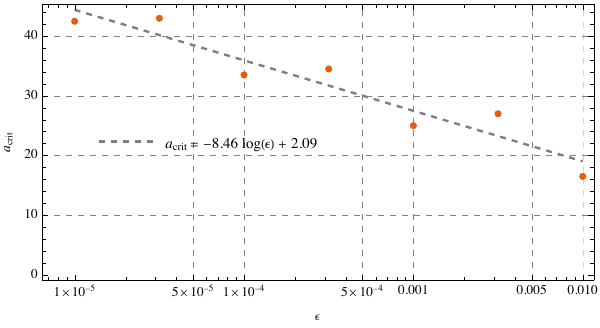}
    \caption{Scaling relation of the critical position $a_{\rm crit}$ as a function of the perturbation amplitude $\epsilon$, shown for the dCS axial mode with $\beta=1$. The orange data points (from Tab.\ \ref{tab:scaling}) track the value of $a_{\rm crit}$ as $\epsilon$ is varied. The gray dashed line represents the linear fit $a_{\rm crit} = -8.46 \log(\epsilon) + 2.09$. As the perturbation $\epsilon$ decreases, the critical point $a_{\rm crit}$ is systematically pushed to larger distances.}
    \label{fig:scaling}
\end{figure}

A linear fit to the data $a_{\rm crit}$ vs. $\log(\epsilon)$ yields a high goodness of fit with $R^2 \simeq 0.91$, as shown in Fig.\ \ref{fig:scaling}.
This approximate logarithmic scaling strongly suggests that mode overtaking is a robust physical phenomenon persisting for smaller perturbations. 
It supports the premise that the use of $\epsilon=10^{-2}$ merely acts as a rescaling tool to bring the critical dynamics into a computationally accessible domain, without altering the underlying physical mechanism.

\bibliographystyle{apsrev4-1}
\bibliography{main}

\end{document}